\documentclass[10pt, journal, compsocconf]{IEEEtran}
\UseRawInputEncoding  

\usepackage{url}

\ifCLASSINFOpdf
   \usepackage[pdftex]{graphicx}
\else
\fi

\usepackage{tabu}
\usepackage{booktabs}
\usepackage[tight,footnotesize]{subfigure}
\begin{document}
%

%
%
%
%
\newpage
\onecolumn
\begin{table*}
	\resizebox{0.9\linewidth}{!}{%
		\begin{tabular}{l}		
			\textbf{Copyright © 2023 IEEE}	\\
			\textbf{© 2023 IEEE. Personal use of this material is permitted. Permission from}\\
			\textbf{IEEE must be obtained for all other uses, in any current or future media,}\\
			\textbf{including reprinting/republishing this material for advertising or promotional}	\\
			\textbf{purposes, creating new collective works, for resale or redistribution to servers}\\
			\textbf{or lists, or reuse of any copyrighted component of this work in other works.}\\
			\textbf{}\\  	[1ex] 	           
	\end{tabular}}
	
\end{table*}

\newpage
\twocolumn

\title{Towards Confidential Computing: A Secure Cloud Architecture for Big Data Analytics and AI}
%
%
%
\author{\IEEEauthorblockN{Naweiluo Zhou, Florent Dufour, Vinzent Bode, Peter Zinterhof, Nicolay J Hammer, Dieter Kranzlm\"uller\\}
	\IEEEauthorblockA{ Leibniz Supercomputing Centre (LRZ), Munich, Germany\\
		Email: naweiluo.zhou@ieee.org, Firstname.Lastname@lrz.de }
}

%
%

\markboth{This is the authors' version}%
{Shell \MakeLowercase{\textit{et al.}}: Bare Demo of IEEEtran.cls for Journals}
%



\maketitle

\begin{abstract}

Cloud computing provisions computer resources at a cost-effective way based on demand. Therefore it has become a viable solution for big data analytics and artificial intelligence which have been widely adopted in various domain science. Data security in certain fields such as biomedical research remains a major concern when moving their workflows to cloud, because cloud environments are generally outsourced which are more exposed to risks. We present a secure cloud architecture and describes how it enables workflow packaging and scheduling while keeping its data, logic and computation secure in transit, in use and at rest.

\end{abstract}	


%
\IEEEpeerreviewmaketitle

\section{Introduction}

\IEEEPARstart{C}loud computing offers on-demand compute and storage resources at a cost-effective way, and therefore has become a viable solution for big data analytics and Artificial Intelligence (AI). The two technologies have been applied in various domain science to process massive amounts of data generated by scientific experiments and simulations, and to facilitate decision making. In biomedical research, for instance, AI is utilised to detect and predict diseases, and improve personalised medicine \cite{HARVEY2012625}. Concern over data security is raised when moving such applications to the cloud because of data sensitivity in these fields. Data anonymisation is a typical way to preserve data privacy, however, certain information such as genome that is unique to any individual, requires further protection strategies against threats in cloud.

Often, hardware resources are shared among users in a cloud environment so as to achieve cost efficiency and maximise resource usage, \textit{e.g.} multiple virtual machines (VM) share one physical host. Virtualisation not only offers resource efficiency, but also preserves isolation at software level. Resource isolation reduces likelihood of a breach and limits the scope of damage when a breach occurs \cite{10.1145/3365199}. Unfortunately, virtualisation also introduces new risks. For example, VMs are normally managed by hypervisors, and a compromised hypervisor can seize control of all the VMs and accesses their data. Data encryption plays a crucial part in preserving workflow security, while the other imperative part is to encrypt workflow logic and computation \cite{10.1145/3332301}.  Conventionally, workflow encryption deals with its status \textit{at rest} (on disk) and \textit{in transit} (transfer between devices). Confidential computing \cite{9604800} addresses security issues of data and code in use (\textit{e.g. in memory and register}). The above three technologies enable a workflow moving freely from device to device by being encrypted, and only temporarily decrypted for use within an isolate before being encrypted again for store \cite{9604800}.

In this paper, we prototype a secure cloud system that finds a trade-off in terms of security, performance, maintenance and usability. This cloud architecture can safeguard user workflows in transit, in use and at rest. Furthermore, it offers the choice of different security levels depending on the data sensitivity. Three main encryption technologies are adopted to achieve this goal: encrypted containers, encrypted VMs and encrypted storage system in addition to resource isolation. The rest of the paper is organised as follows. Firstly, Section~\ref{sec:related_work} briefly reviews related work. Next, the proposed architecture is described in Section~\ref{sec:architecture}. Followed, some preliminary results are given in Section~\ref{sec:resulst}. Lastly, Section~\ref{sec:future_work} concludes the paper and proposes future work.

\section{Related Work}\label{sec:related_work}

H\"ob \textit{et al.} \cite{H_b_2020} presented a software framework that enables automatic converting of Docker \cite{ Merkel2014} container images into Charliecloud \cite{10.1145/3126908.3126925} solutions; and scheduling thereof onto high performance computing (HPC) systems. CharlieCloud is a container engine specifically designed for HPC systems. We adopt the concept of automatic container image generation and scheduling. Additionally, we include container encryption and adapt the scheduling strategies for Cloud systems.

Nolte \textit{et al.} \cite{9826008} described a secure workflow targeted for HPC systems aiming to process sensitive data on a system that presumes to be untrusted. Workflow security is realised by isolating compute nodes; encapsulating workflow inside encrypted containers; and requiring decryption keys to mount data into the workflow containers. Asymmetric encryption and decryption keys are generated and distributed to users via a key management system Vault\footnote{\url{https://github.com/hashicorp/vault} (accessed on 17/03/2023)}. Similarly, our architecture adopts the idea of container encryption and makes use of Vault to ascertain application security.

\begin{figure*}[!ht]
	\centering
	
	\includegraphics[width=.65\textwidth]{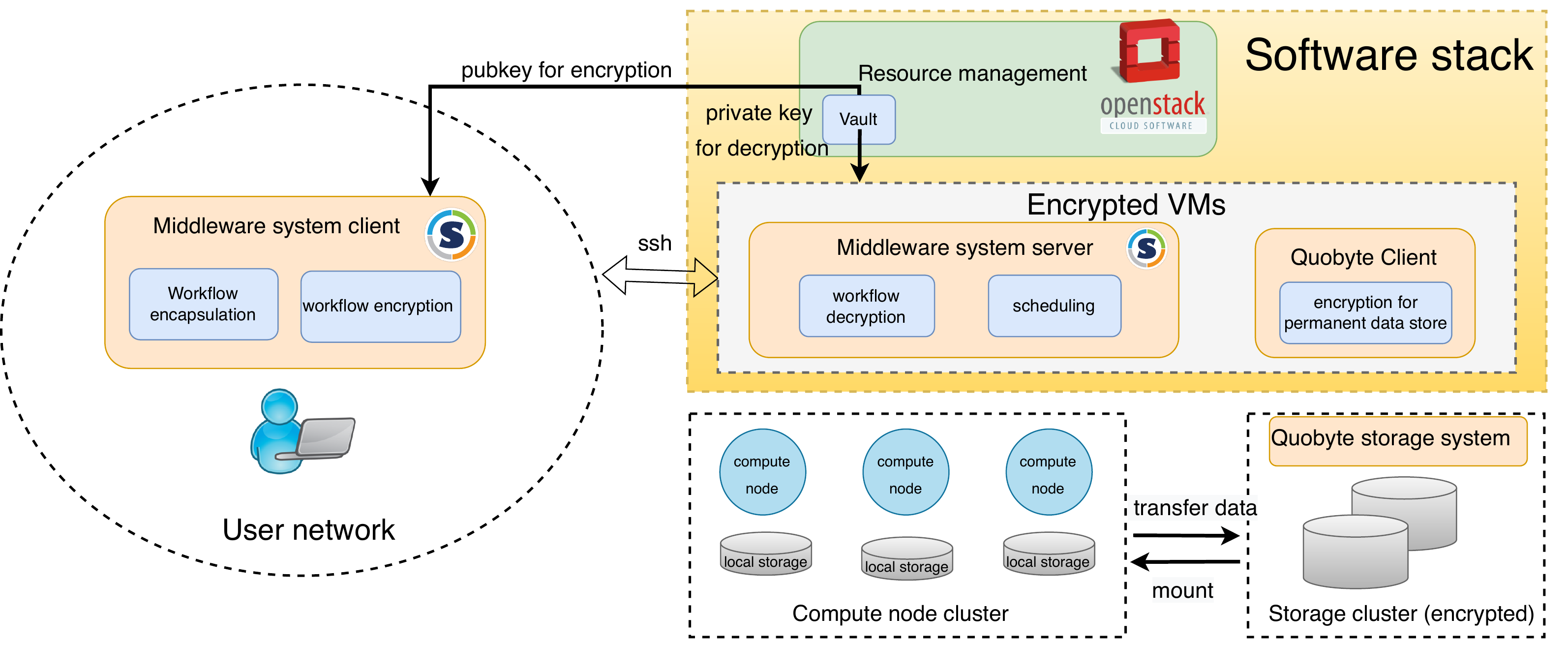}
	\caption[]{Architecture of the secure cloud powered by confidential computing technologies. }
	\label{fig:arch_testbed}
\end{figure*}
\section{Design of the Secure Cloud}\label{sec:architecture}
This section describes our cloud architecture as illustrated in Fig.~\ref{fig:arch_testbed} including hardware and software stack. Table~\ref{table:softwarestack_list} lists the main software. The hardware incorporates a compute node cluster, storage located within the same network as the compute nodes (referred as local storage herein, limited capacity), and storage (large capacity) located in a separate network. The software stack performs three main functions: resource provision, threat defence, and program encapsulation and scheduling.

\begin{table}[ht]
	\centering  
		\resizebox{0.9\linewidth}{!}{%
	\tabulinesep=0.6mm
	\begin{tabu}to 0.5\textwidth{ l| X} 
		\toprule	
		Software name & Description\\
		\bottomrule
		OpenStack & Resource provision \\
		
		AMD SEV & VM encryption for data in use\\
		
	Middleware system & Containerisation, encryption and scheduling\\
	Singularity& Containerisation, encryption \\
		Vault & Key management system \\
		
		Quobyte& Storage management system; data encryption at rest \\
		\bottomrule	
	\end{tabu}	
}
	\caption[]{The software stack of the Cloud architecture. }\label{table:softwarestack_list}
\end{table}

\subsection{Resource Provision}

Our architecture utilises OpenStack to (1)provision the VM resources and local storage; (2)offer network isolation; and (3)enforce role-based access control and organisation-based access control. 

\subsection{Workflow Containerisation And Scheduling}
The Middleware system as illustrated in Fig.~\ref{fig:arch_testbed} is composed of two parts: the \textit{middleware system client }and the \textit{middleware system server}. The \textit{client} is distributed to users for encapsulating their workflow inside a Singularity \cite{Kurtzer2017SingularitySC} container image. Containers enable application portability and environment compatibility. In case of big data sets, it is packed in a separate LUKS (Linux Unified Key Setup) filesystem image. The \textit{server} schedules the containerised workflow to multiple VM nodes via invocation of MPI (Message Passing Interface) \cite{mpistandardv3.1}, and passes the private key to decrypt the containers and LUKS images.

\subsection{Security Measures To Defend Threats}\label{threat_model}

This section describes three types of data encryption strategies. Furthermore, network separation, storage partition and isolation (\ref{encryption_at_rest}) are proposed to protect data at rest.
\subsubsection{Workflow Encryption In Transit}\label{subsubsec:in_transit}
Besides \texttt{ssh}, which is the \textit{de facto} standard secure way to connect remote machines, workflows are encrypted during transmission from a user network to the cloud to ward off network security breaches. The \textit{middleware system client} builds an encrypted Singularity image. The encryption and decryption are performed asymmetrically via a RSA key pair (\textit{i.e.} pubkey and private key). The pubkey is generated and offered to users via a key management system Vault. Similarly, LUKS images are encrypted and decrypted using the same key pair.

\subsubsection{Encryption In Use}\label{subsubsec:encry_in_use}

Container decryption only occurs shortly at runtime within kernel, and the container image stays encrypted when stored on disk. Vault passes the decryption key to the user's VMs. This key only resides on the encrypted storage system mounted to the VM nodes (see section~\ref{encryption_at_rest}).  When a higher security level is required, which is likely to bring performance vicissitude, users can choose to spawn the VM nodes powered by the AMD SEV (Secure Encrypted Virtualisation) technology \cite{amd2020}. AMD SEV is a pioneer technology towards confidential computing through memory encryption, which isolates a hypervisor from its VMs. When a hypervisor reads the VM memory, it only sees encrypted bytes. This protects VMs from malicious administrators and defends attacks to VMs from a compromised hypervisor.

\subsubsection{Encryption At Rest}\label{encryption_at_rest}
The storage  is split into two isolated networks for security reasons. The local store is a general filesystem where a user's workflow at rest is encrypted inside a Singularity image and a LUKS image. The data for long-term and more secure store can be moved to the Quobyte storage system where the disk is encrypted with symmetric AES-XTS  \cite{ieeextsstandard} algorithm using a 128-bit or 256-bit cryptographic key. Quobyte\footnote{\url{https://www.quobyte.com/} (accessed on 13/03/2023)} also partitions the storage into isolated domains, which enables organisation-wise accesses restrictions. Without Quobyte, user workflows experience reduced security.

\section{Preliminary Performance Analysis}\label{sec:resulst}

The preliminary results show performance costs brought by the two encryption technologies in our architecture: Quobyte system encryption and Singularity encryption. To evaluate encryption cost introduced by Quobyte, we measure the I/O bandwidth fluctuations using the IOR benchmark\footnote{\url{https://ior.readthedocs.io/en/latest/intro.html} (accessed on 13/03/2023)}.  The execution time difference caused by Singularity encryption is evaluated with an MPI benchmark: BPMF \cite{ conf/cluster/AaCH16}. The BPMF benchmarks are containerised inside Singularity.

\begin{figure}[!ht]
	\centering 

		\includegraphics[width=0.4\textwidth]{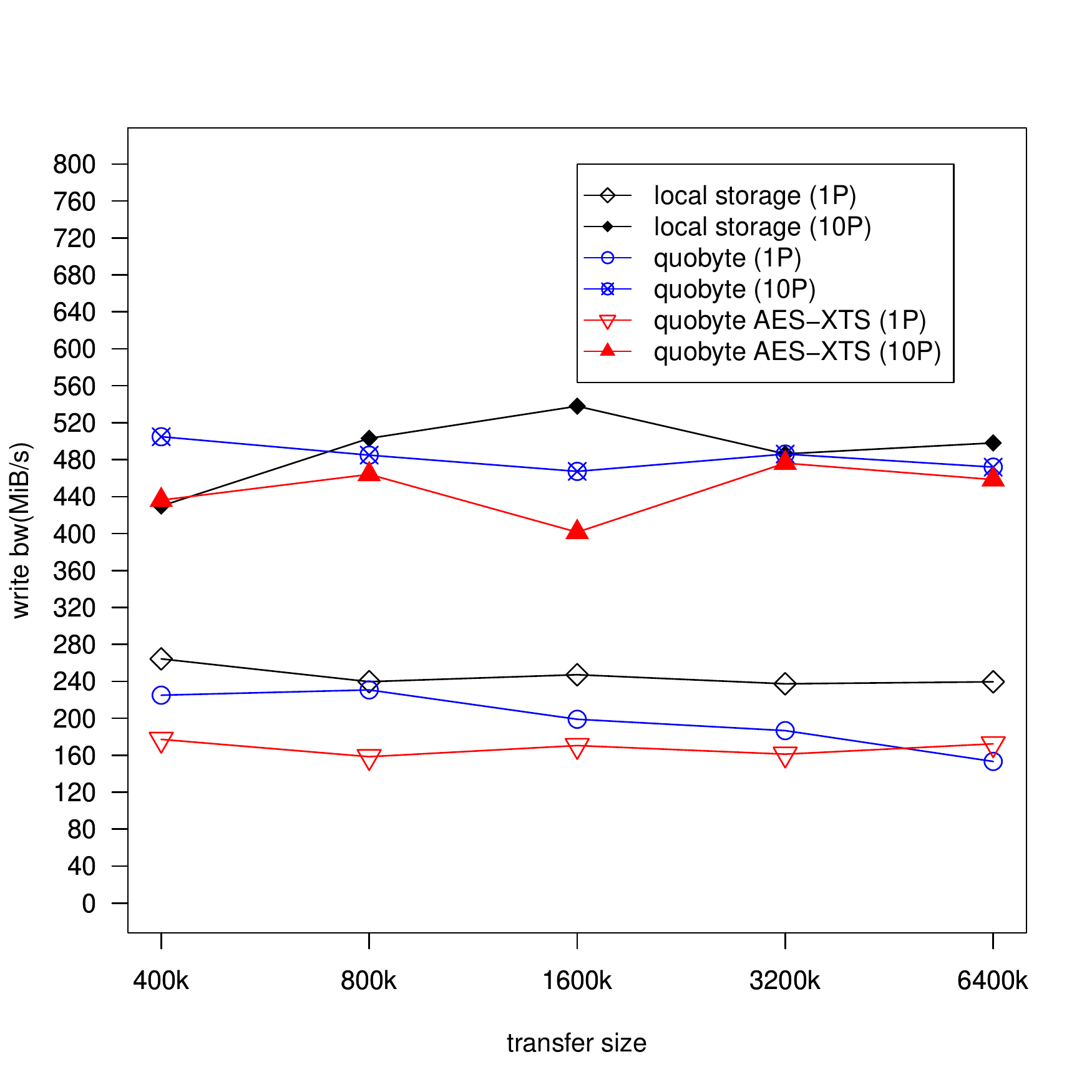}
	
	\caption[]{The I/O bandwidth comparison for the normal local filesystem, Quobyte storage system (with and without 128-bit AES-XTS encryption). The bandwidth shows the operations of transferring various sizes of data blocks with 1 process and 10 processes. Local storage refers to the storage within the same network as the compute node. }
	\label{fig:io_cost}
\end{figure}

\begin{figure}[!t]
	\centering
	\includegraphics[width=0.45\textwidth]{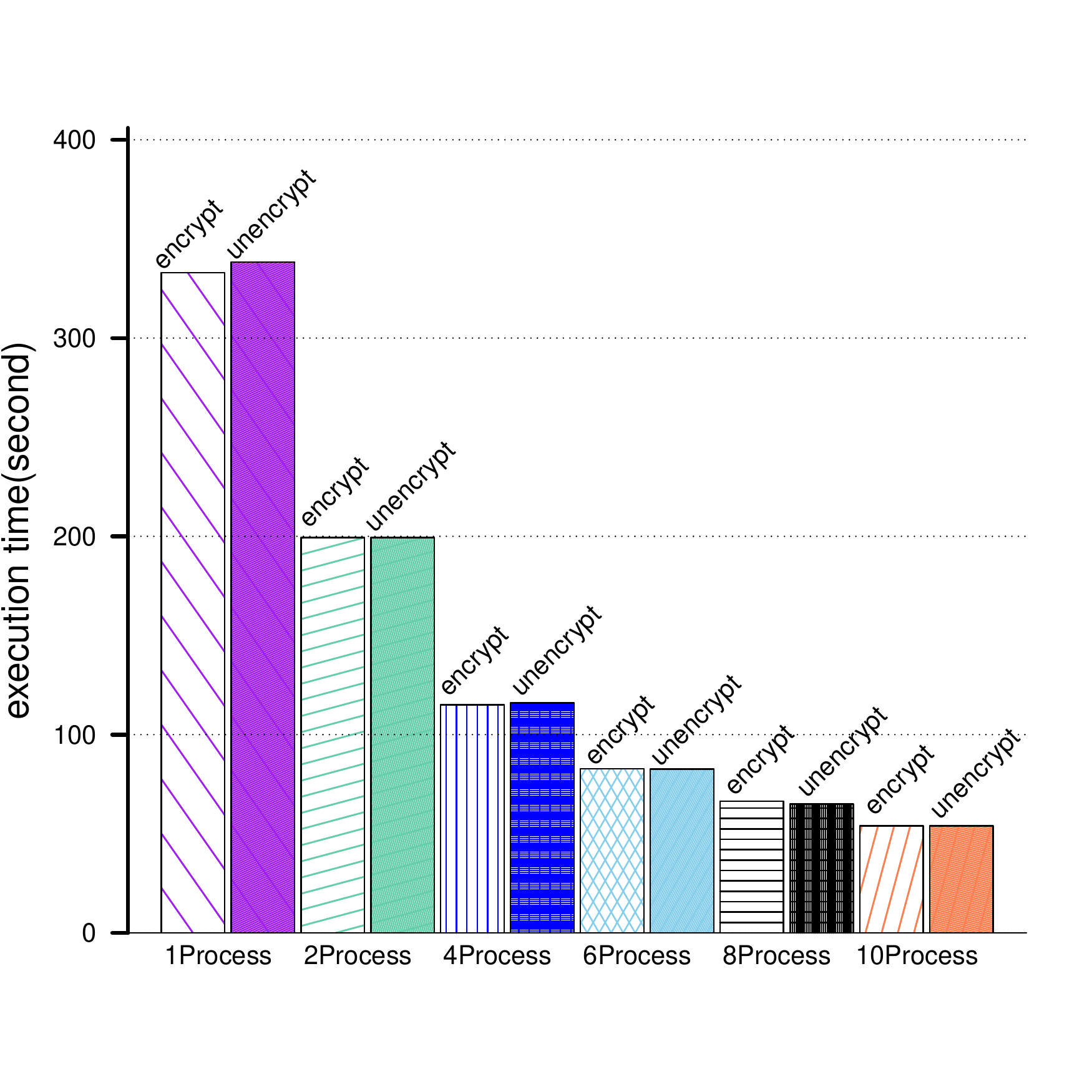}
	\caption[]{Execution time of BPMF encapsulated in encrypted and unencrypted containers running with different process numbers on one single node. }
		
	\label{fig:singularity_cost}
\end{figure}

Fig.~\ref{fig:io_cost} gives the comparison on I/O bandwidth for the local storage and Quobyte-managed storage (\textit{i.e.} encryption with 128-bit AES-XTS and encryption disabled). We herein only show the speed of write operations. The local storage shows a higher bandwidth comparing with Quobyte-managed storage, however, its magnitude is insignificant. In contrast to \textit{a priori} reasons, the performance impact caused by Quobyte encryption is negligible. This may be emerge from network latency that costs more than encryption. Fig.~\ref{fig:singularity_cost} presents the execution time of BPMF with and without Singularity encryption. No visible performance degradation is introduced by container encryption.

\section{Conclusion and Future Work}\label{sec:future_work}
This article presented a secure cloud architecture that safeguards user workflows and enables moving them freely between user and cloud. A salient feature of this cloud system is that it offers the flexibility to provision resources as a general purpose cloud or a highly-secured cloud to host domain science workflows that have different data sensitivity.

Next,  VM encryption cost will be benchmarked, which is estimated to give a non-trivial impact on performance. To alleviate the performance decline, future work will focus on scalability of compute resources and introduce GPU supports. Performance evaluation will be carried out to compare scheduling efficiency of applications in big data analytics and AI when different security levels are featured.

\section*{Acknowledgment}
We kindly acknowledge the support of  Bavarian State Ministry of Health and Care who funded this work with DigiMed Bayern (Grant No: DMB-1805-0001) within its Masterplan “Bayern Digital II”.


%

%
%
%
%
%

\ifCLASSOPTIONcaptionsoff
  \newpage
\fi

\end{document}